\documentclass[useAMS,usenatbib,usegraphicx]{mn2e}

\def\kms{$\rm km\, s^{-1}$}
\def\cm3{$\rm cm^{-3}$}
\def\Ts{$\rm T_{*}$}
\def\Vs{$\rm V_{s}$}
\def\n0{$\rm n_{0}$}
\def\B0{$\rm B_{0}$}

\def\erg{$\rm erg\, cm^{-2}\, s^{-1}$}
\def\Hb{H$\beta$}

\title[R Aquarii spectra revisited by SUMA]{R Aquarii spectra revisited by SUMA}
\author[M. Contini and L. Formiggini]{Marcella Contini$^{1}$\thanks{E-mail:
contini@ccsg.tau.ac.il} and Liliana Formiggini$^2$ \\
$^{1}$School of Physics and Astronomy, Tel Aviv University, Tel Aviv 69978, Israel\\
$^{2}$ School of Physics and Astronomy, Tel Aviv University and the Wise Observatory, Tel Aviv 69978, Israel}
\begin{document}

\date{Accepted. Received ; in original form }

\pagerange{\pageref{firstpage}--\pageref{lastpage}} \pubyear{2002}

\maketitle

\label{firstpage}

\begin{abstract}

We analyse the optical spectra and   the UV spectral evolution  of the jets
and  of the HII region inside the R Aquarii binary system by the code SUMA
which consistently accounts for shock and photoionization.
The temperature of the hot star results 80,000 K as for a white dwarf.
We find that the shock velocity in the NE jet increased  between 1983  and 1989.
The spectral evolution between 1989 and 1991  of the SW jet indicates that
a larger  contribution from low density-velocity matter  affects the  1991 spectra.
The evolution of the UV spectra from 8/11/1980 to 26/5/1991 in the HII region
indicates that the reverse shock is actually a standing shock.
The results obtained by modelling the line spectra are cross-checked by the fit
of the continuum SED. It is found that a black-body temperature of 2800 K
reproduces the radiation from the red giant. A  black-body emission component
corresponding to 1000 K  is emitted by dust in the surrounding of the red giant.
Model calculations confirm that the radio emission is of thermal origin.
We found that the NE jet bulk emission is at a distance  of
about 2 10$^{15}$ cm from the internal system, while the distance of the SW jet bulk is
$\sim$ 6 10$^{14}$ cm.
The distance of the reverse shock from the hot
source in the internal region is  $\leq$ 9 10$^{13}$ cm.

\end{abstract}


\section{Introduction}

The symbiotic star  R Aquarii  (R Aqr) is an interacting system of a long-period
Mira variable (LPV) and a hot star.
The optical light curve is dominated by the  386.83 day pulsation period of the LPV, while
the presence of a hot companion and its accretion disk is inferred
from the ultraviolet  spectra.
The binary system is characterized by
 a highly inclined orbit to the line of sight
(i$\sim$ 70 degree) with large eccentricity (e $\sim$0.8), a
semi-major axis of 2.7 10$^{14}$ cm  (Meier \& Kafatos 1995), and
an orbital period of  $\sim$ 44 yr (Wilson et al 1981, Hinkle at al. 1989,
Hollis, Pedelty, \& Lyon 1997).  A binary separation of 11 AU is evaluated
adopting a  distance  to Earth   d $\sim$ 200 pc (Solf \& Ulrich 1985, Hollis et al. 1997;
see also Viotti et. al 1997).

The  binary system  hosts a compact HII region within
a filamentary oval nebula (Kafatos, Michalitsianos, \& Hollis 1986)
and is surrounded by a  large and complex nebulosity.
Two jets oriented at SW and NE have been detected and  were observed at
ultraviolet, optical, and radio wavelengths.  The bipolar
morphology of R Aqr detected by VLA data (Kafatos et al 1989) was
confirmed by HST high resolution images (Paresce et al 1991, Paresce
\& Hack 1994).
Meier \& Kafatos (1995, Fig. 1) show the position of the NE and SW jets with
respect to the central region of R Aqr.

The extensive IUE monitoring of the  spectra from the central HII region and from the NE
and SW jets has been analysed by Hollis et al (1991) and Meir \&
Kafatos (1995).
In particular, the HII region and the NE jet temporal changes show
increasing emission line intensities from 1979 to 1989, while
the SW jet shows  a decrease by a factor of 1.5  between 1989 and 1991.

Both shocks  and photoionization by the hot star
have been evoked as the excitation mechanisms that produce the
rich emission line spectra  from the jets (Hollis et al 1991).
The observed line ratios  could not be
explained by pure photoionization models and  also shock dominated models
(Binette et al 1985) were not successful to reproduce them.
Some discrepancies also result  from the modelling of Burgarella, Vogel, \& Paresce (1992)
who accounted for shock and photoionization by the hot source,
which were not calculated in a consistent way.

In this paper we would like to  investigate the R Aqr system adopting the colliding
wind scenario.
In symbiotic binaries the collision of the  winds from the hot star  and  the red giant
 creates a complex hydrodynamic structure (Nussbaumer 2000 and references therein).
The  emitting gas within the system is then ionized and heated both by
 the photoionizing flux from the hot star and by shocks.

We will follow the same procedure as
previously adopted  for HM Sge (Formiggini, Contini, \& Leibowitz 1995),
AG Peg (Contini 1997), RR Tel (Contini \& Formiggini 1999), etc., focusing on
the physical conditions in the emitting nebulae
by  modelling  the emission spectra in the central HII region and in the jets.

Models which consistently account for shock and photoionization from an external source
are used (Sect. 2),  adopting for the calculations the code SUMA
(see Viegas \& Contini 1994, Contini 1997, and references therein).

We will start  by modelling the optical and UV line ratios in the SW  and NE jets (Sect. 3).
We will investigate  the time evolution of the UV spectra from  the HII region in Sect. 4,
since no optical spectra are given for the internal region of R Aqr.
The models will be cross checked by the agreement with the spectral energy distribution
(SED) of the continuum (Sect. 5). Concluding remarks appear in Sect. 6

\section[]{The models}

\begin{figure}
\includegraphics[width=84mm]{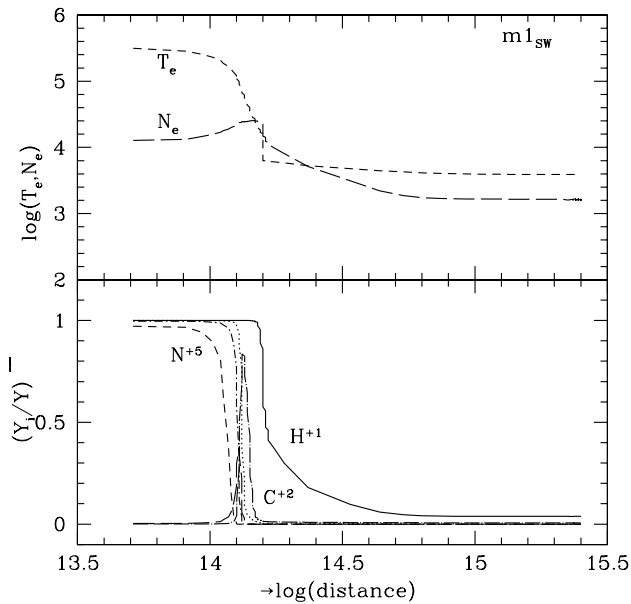}
\caption
{The shock front is on the left.
bottom: The distribution of the fractional abundance throughout
the nebula relative to model m1$_{SW}$ of the ions  N$^{+5}$
(short-dashed line), C$^{+4}$ (dot-dashed line), C$^{+3}$ (long-dashed line),
C$^{+2}$ (long-dash-dot line), H$^+$ (solid line), He$^{+2}$ (dotted line);
top: the distribution of the electron temperature and electron density.}
\label{Fig.1}
\end{figure}

\begin{figure}
\includegraphics[width=84mm]{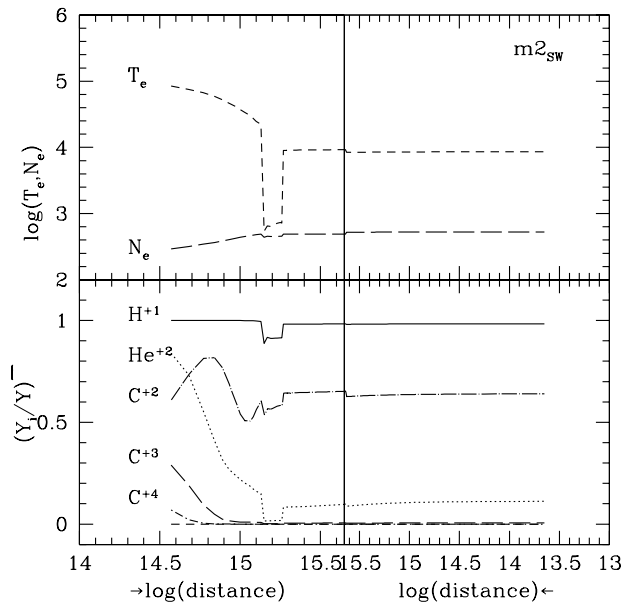}
\caption
{The same as for Fig. 1 for model m2$_{SW}$.
The  shock front is on the left and radiation reaches the right edge.
The diagram is divided in two equal symmetric  parts in order to give 
equal importance to the shock dominated region (left) and to the
radiation dominated one (right).}
\end{figure}

\begin{figure}
\includegraphics[width=84mm]{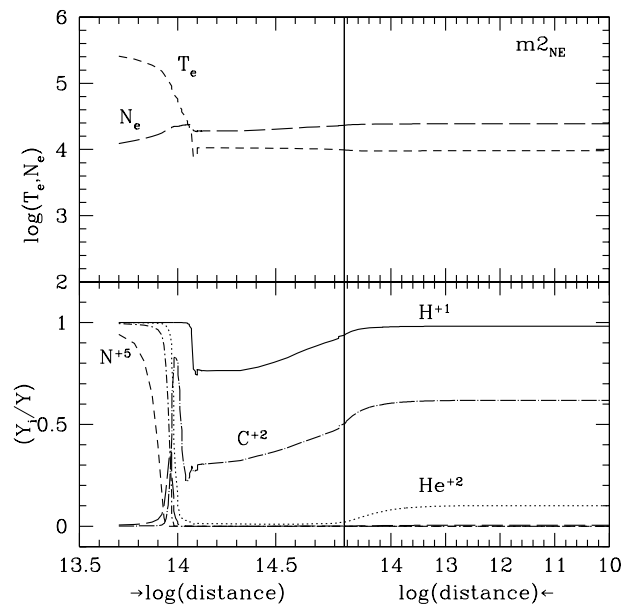}
\caption
{The same as for Fig. 2 for model m2$_{NE}$.}
\end{figure}

\begin{figure}
\includegraphics[width=84mm]{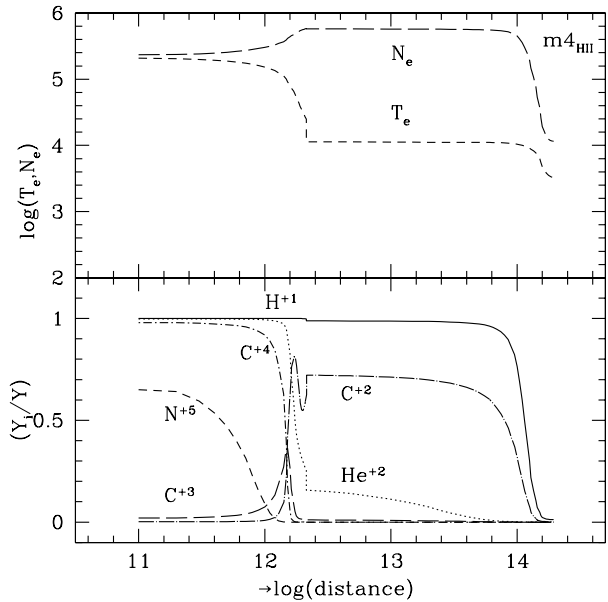}
\caption
{The same as for Fig. 1 for model m4$_{HII}$.}
\end{figure}

Two shock fronts are created at collision
of the winds between the  stars, one propagating in reverse
toward the  hot star and  the other propagating outwards the system.
Recent observations of symbiotic novae show that spectra are also
emitted from the downstream region of shocks propagating outside
the system and eventually merging with the ISM (Contini \& Formiggini 2001).

The  nebula between the stars
is ionised and heated by both black body radiation from the  hot star and  the (reverse) shock
(see Contini \& Formiggini, 2000, Fig. 1).
Therefore, radiation reaches the very  shock front edge of the nebula.
On the other hand, radiation and shock act on opposite edges of the nebula
downstream of the expanding shock which propagates in the red giant atmosphere.

SUMA calculates the emitted spectrum from a gas in a  plane parallel geometry.
The model input parameters are those which describe the shock
(the shock velocity, \Vs, the preshock density, \n0, and the
preshock magnetic field, \B0), those which represent the radiation flux
(the colour temperature of the hot star, \Ts, and the ionization
parameter, U), the geometrical thickness of the emitting
nebulae, D, the dust-to-gas ratio, d/g, and the relative
abundance to H of the elements, He, C, N, O, Ne, Mg, Si, S, A, and Fe.

The choice of the  input parameter ranges  is described in the following.

The radiation flux from the hot star is a black body distribution of
temperature \Ts  ~selected by the best fit of the line ratios (NV/CIV, HeII/\Hb, CIII]/CIV, etc.).
 \Ts  was evaluated to  40 000 K by Bulgarella et al (1992) leading, however,
to underpredicted  HeII lines, and between 50 000 K and 60 000 K by Meier \& Kafatos (1995)
by the Zanstra method.
They claim that  the NV excitation would need a higher \Ts, up to 80,000 K,
and  they also claim that "the overall ionization state obtained by IUE
does not provide evidence for such high state of ionization".
However, the ionization structure  of the emitting nebulae is not easily predicted.
We will select \Ts ~by the best fit of the line ratios (Sect. 3.1.1).

We  adopt \B0 between 10$^{-3}$ - 10$^{-4}$ gauss,  that was found for other symbiotic systems.
The relatively high magnetic field is in agreement with   the magnetic field
in the environments of isolated giants (Bohigas et al. 1989).

\Vs ~is  indicated by the FWHM of the line profiles, although the shock
velocity is not exactly the velocity of the gas.
Hollis et al (1991, Fig. 3)   suggest velocities of $\sim$ 150 \kms in the jets and of $\sim$ 90 \kms
in the HII region. However, Michalitsianos, Perez, \& Kafatos (1994)
give evidence to velocities up to 300 \kms
(see Sect. 5).

\n0 is selected from the ratio of typical lines. Notice, however,
that the lines are emitted from the gas downstream, where compression leads
to a  density  $>$ \n0 by  a factor which depends on \Vs, \B0, and \n0.

The geometrical thickness of the emitting nebula, D, is determined phenomenologically by the best fitting
models but is constrained by the dimensions of the system.

We adopt relative abundances close to  solar
(He/H=0.1, C/H=3.3 10$^{-4}$, N/H= 10$^{-4}$, O/H=5. 10$^{-4}$, Ne/H=8.3 10$^{-5}$,
Mg/H=2. 10$^{-5}$, Si/H= 3.3 10$^{-5}$, S/H=1.6 10$^{-5}$, Ar/H=6.3 10$^{-6}$, and
Fe/H=4. 10$^{-5}$)
as a first guess, refining them by the
fitting process of the spectra emitted from different regions throughout
the system.

A dust-to-gas ratio by mass, d/g = 4. 10$^{-4}$  is used for all models.

To obtain the best fit of  calculated to  observed  data the spectra emitted
from the different nebulae within the R Aquarii system are summed up
adopting relative weights. The  relative weight, w,  is proportional to the
emitting   area of  the nebula corresponding to   a model.

The models are described in Table 1.

For sake of clarity in the following, models referring to single nebulae
are indicated with  m (e.g. m1$_{SW}$); models referring to
averaged results are indicated by the capital M (e.g. M1$_{NE}$).

\section{Line spectra  from the jets}

\begin{table*}
\centerline{Table 1}
\centerline{The models adopted in Table 2, Table 3, and Table 4}
\begin{tabular}{ ll l l lllll } \\ \hline
    &  m1$_{SW}$&m2$_{SW}$&m1$_{NE}$&m2$_{NE}$ &m3$_{NE}$& m4$_{NE}$ \\ \hline
\ \Hb  ~(\erg)      & 3.7(-3)& 3.2(-4) &0.122&0.078&0.08&4.(-4)  \\ 
\   \Vs ~(\kms)  & 150. &80.& 120.& 140. &150.& 50. \\
\   \n0 ~(\cm3)  &   4(3) & 100.&7.(3)&4.(3) &4.(3)& 50.    \\
\   \B0 ~(gauss) &   1(-3) & 1.(-4)&1.(-3)&1.(-3)&1.(-3)& 1.(-4)  \\
\   D (10$^{15}$ cm)  & 2.5& 9.&1.4& 1.4&1.4& 10.   \\
\   U            &  - &2.3(-3)&1.6(-3)& 2.2(-3) &2.2(-3)& 1.(-3)  \\ \hline
&&&&&&\\

\end{tabular}
\end{table*}

In the present investigation we will  treat the optical and UV spectra separately,
because of the different sizes of the entrance apertures (see Hollis et al. 1992).
The observations in the optical range  by
Hollis et al (1991) have been obtained in November-December 1989.
In the UV we  refer to  Meier \& Kafatos (1995) who analysed
a set of good signal-to-noise low resolution IUE spectra of R Aqr, between 1979
and 1992.
The compact HII region and the NE jet were observed in several
epochs, while for the SW jet only two spectra are available.
The line intensity errors  were evaluated to 4 \% - 5\% for the strongest
lines and up to 10 \% - 15 \% for the moderately exposed ones.

The input parameters suggested by Hollis et al. (1991) are taken as a first
choice; they are, however,  modified in order to achieve the best
agreement with the data.

\subsection{The SW jet}

First we consider the SW jet, which corresponds to the A' feature
in Meier \& Kafatos (1995, Fig. 1, bottom).
 The aperture of the
spectrograph  was carefully positioned to receive  only the  jet contribution, and the data
by Meier \& Kafatos (1995, Table 2C)  represent the emission solely from
the SW jet (S. R. Meier,  private communication).

\subsubsection{The optical line ratios}

\begin{table*}
\centerline{Table 2}
\centerline{Optical line intensities relative to  \Hb =1 in the jets}
\begin{tabular}{llllll ll} \\ \hline  
 line  & obs(SW)$^1$& m1$_{SW}$ & m2$_{SW}$ & M$_{SW}$(opt)$^2$& obs(NE)$^1$ & m2$_{NE}$&m3$_{NE}$ \\
\hline
\ [OII] 3727+ & 10.11 & 10.7 & 5.0  &9.53  & 2.3 & 2.66 &3.0\\
\ [NeIII] 3869+ & 2.68 & 2.6 & 1.0  & 2.4 & 0.85 & 1.1 &1.3\\
\ [SII] 4069+  &0.33 & 1.1 & 0.04 & 0.70  & 0.44 & 0.7 &1.\\
\ [OIII] 4363 & 0.67 & 1.3 & 0.27 & 1.0  & 0.14 & 0.10 &0.1\\
\ HeI 4471     & $<$ 0.28 & 0.17 & 0.06  & 0.15 & 0.08 & 0.05 &0.06\\
\ [OIII] 5007+ & 15.14 & 15.6& 8.2  & 14.1 & 4.2  & 4.05 &4.2\\
\ [NI] 5200+   & $<$ 0.28 & 0.18 & 0.01 & 0.15 & 0.09 & 1.(-5) &1.(-5)\\
\ [NII] 5755   & 0.74 & 0.4 & 0.04 & 0.33  & 0.09 & 0.09&0.1 \\
\ HeI 5876    & 0.80 & 0.53 & 0.20 & 0.5   & 0.12 & 0.15 &0.16\\
\ [OI] 6300+  & $<$ 0.8 & 0.9 & 0.05 & 0.72  & 0.62 & 0.23 &0.4\\
\ [NII] 6548+  & 5.45 & 6. & 2.19& 5.2 & 3.17 & 4.6  &5.0\\
\ [SII] 6717   &0.73 & 0.8 & 0.27 & 0.7 & 0.25 & 0.22&0.3  \\
\ [SII] 6730  & 0.63& 1.3 & 0.26 & 1.0  & 0.41 & 0.49 &0.7 \\
\ [OII] 7322+  & $<$ 0.28 & 6.7 & 0.23& 5.3 & 0.87 & 0.8 &1.1\\
\ [SIII] 9069+ & $<$ 0.56& 1.17 & 2.66& 1.47 & 1.34 & 2.4 &1.6\\
\ CIV 1550      & - & 91.7 &  5.56 &- &  -   & -&-\\
\ w$^3$              &-& 1. & 3.   & - & - & - & - \\ \hline\\
\end{tabular}

\flushleft

$^1$ Hollis et al. (1991, Table 3)

$^2$ The averaged line ratio  is :

I$_{\lambda}$/I$_{H\beta}$= ((I$_{\lambda}$/I$_{H\beta}$)$_1$ \Hb$_1$ w$_1$ +
(I$_{\lambda}$/I$_{H\beta}$)$_2$ \Hb$_2$ w$_2$) / (\Hb$_1$ w$_1$ + \Hb$_2$ w$_2$)

$^3$ relative weight

\end{table*}

In Table 2 the optical observed line intensity ratios to \Hb ~(column 2)
are compared with model results (columns  3, 4, and 5).
Although the high [OII]/\Hb ~and [OIII] 5007+/\Hb ~line ratios suggest that a shock
dominated (SD)
model   should explain the optical spectrum, the [SII] 6717/6730 line ratio ($>$ 1)
is not well fitted by model m1$_{SW}$, which is characterized by U=0 and a high \n0.

So, another component must be considered with  lower \n0 (and \Vs).
We present in Table 1 the radiation dominated (RD) model m2$_{SW}$ calculated with
\n0 = 100 \cm3, \Vs = 80 \kms and \Ts = 80 000 K.
The weighted sum of the two models, M$_{SW}$(opt), is given in Table 2, column 5.
Model m2$_{SW}$ shows some discrepancies  for most of the line ratios,
but a better fit  to  [SII] 6717/6730, and most importantly, it leads to
a better agreement of the summed spectrum  with the data (Table 2).
An RD model, corresponding to \n0=65 \cm3,  which provides
[OIII]/\Hb ~and [OII]/\Hb  ~better fitting  the observations
was discarded  when modelling the UV spectra ((Sect. 3.1.2).
In fact, optical and UV spectra are modelled consistently, and  the RD model
is choosen cross-checking the results in the  UV ranges,
until a fine tuning is obtained.
Adopting w(m1$_{SW}$) : w(m2$_{SW}$) :: 1: 3 (last row of Table 2), which
leads to the best fit of the observed optical spectrum,
the [SII] line ratio is still lower than observed in the  averaged spectrum.
In fact, \Hb ~corresponding to m2$_{SW}$ is  by a factor of $\sim$ 10 lower than \Hb
~calculated by m1$_{SW}$.

The colour temperature of the hot star \Ts=80 000 K, which  is adopted to explain the
optical spectrum from the SW jet, was selected from a large grid of models
calculated with different \Ts. This temperature was found too high by Meier \& Kafatos
to match all the UV spectral line fluxes.
However,
 compression  downstream of the shock front leads to high densities which
speed up   the cooling rate and  a region of low temperature gas also appears inside the nebula.

For sake of consistency, this value will be adopted
hereafter to model the spectra from the NE jet and the HII region.

\subsubsection{The  evolution of the UV spectra}

\begin{table*}
\centerline{Table 3}
\centerline{UV line  models (\Hb=1)}
\begin{tabular}{ ll l l l l l ll } \\ \hline
 line & m1$_{SW}$&m2$_{SW}$&m1$_{NE}$& m2$_{NE}$ &m3$_{NE}$& m4$_{NE}$\\
\hline
\ NV 1238+42&38.9&1.9(-3)&1.2 &1.5 &1.67& 5.(-6) \\
\ CII 1334+36 &20.&2.6 &0.93& 0.8&0.87&1.1 \\
\ SiIV + OIV] 1394+  &69.0&2.2&2.6& 2.97&2.9&0.22\\
\ NIV] 1483+86&22.0&0.4&1.1& 0.9&1.0 &7.2(-3) \\
\ CIV 1548+51 & 92.0&5.56&3.9 & 3.87&4. & 0.068 \\
\ HeII 1640 &0.6&0.33 &0.23& 0.3&0.29 & 0.4 \\
\ OIII] 1661+66&10.2&1.62 &0.45& 0.54&0.55& 0.45\\
\ NIII] 1749+54&11.0&1.6  &0.40& 0.5 &0.5& 0.44  \\
\ SiII 1808+17&0.5&0.06  &0.083&0.075 &0.12& 0.05 \\
\ SiIII] 1883+92&16.3&2.4  &0.72& 0.69 &0.76& 1.5\\
\ CIII] 1907+10&46.&9.46 &2.3 & 2.6 &2.85& 2.2 \\
\ [OIII] + CII] 2321+&21.2&2.5 &2.1&1.54&2.1&1.8  \\
\ [OII] 2470&5.8&0.3&0.78&0.67&0.9&0.15\\
\ MgII 2795+&0.23&0.06 &0.38&0.32&0.45&0.73\\
\ HeII 3203&0.037&0.02&0.014& 0.018 &0.018& 0.024  \\
\ w &1&300 &1&1&  1&350\\
\hline\\
\end{tabular}

\flushleft

$^1$ relative weight

\end{table*}

\begin{table*}
\centerline{Table 4}
\centerline{UV line intensities relative to  CIV (SW Jet)}
\begin{center}
\begin{tabular}{ ll l l l } \\ \hline  
 line & 28/12/1989 & M1$_{SW}$ & 27/5/1991 &M2$_{SW}$ \\ \hline
\ NV 1238+42& 0.19&  0.17 & 0.17&0.15\\
\ CII 1334+36& 0.47 & 0.45& 0.5& 0.42\\
\ SiIV 1394 & 0.31  & 0.56&0.3& 0.54\\
\ OIV] 1401+07 & 0.062&$\uparrow$& 0.3&$\uparrow$\\
\ NIV] 1483+86&0.16?& 0.14& 0.16?& 0.13\\
\ CIV 1548+51 &1.&  1.&1.&1.\\
\ HeII 1640&0.1   &  0.077&0.08& 0.08\\
\ OIII] 1661+66 & 0.33 & 0.22& 0.31&0.24\\
\ NIII] 1749+54 & 0.30 & 0.22& 0.39& 0.23\\
\ SiII 1808+17 &0.06 & 0.008& - &0.008  \\
\ SiIII] 1883+92 &0.27 & 0.31& 0.48& 0.32\\
\ CIII] 1907+10 & $>$ 1.28 & 1.20& $>$ 1.47& 1.24\\
\ w(SD): w(RD)  & - &1:300& - & 1:350  \\ \hline\\

\end{tabular}

\end{center}

\end{table*}

Model results for the SW jet are given in Table 3 and are compared with the data in Table 4.

The line ratios are given relative to CIV, which is generally a strong line.
However,  high   CIII]/CIV and CII]/CIV  ($>$ 1) characterize the
R Aqr UV line spectra.
The  spectra which  appear in columns 3 and 5 of Table 4 (M1$_{SW}$ and M2$_{SW}$,
respectively)
represent the weighted sums of the two models, m1$_{SW}$ and m2$_{SW}$,
 that  were selected by  fitting  the optical lines  (Table 1).
The summed spectra
in the two epochs correspond to different relative weights of the SD and RD models.
Adopting, on the other hand, the relative weights indicated by the best fit of the optical lines
we could not obtain an acceptable  agreement  to  the observed UV line ratios.
Different relative weights (w(SD) : w(RD) :: 1 : 300 in 1989 and w(SD) : w(RD) :: 1: 350 in 1991,
instead  of 1 : 3 as in the optical range) lead to the
best fit of the UV spectra.
Although the fit is  rough the trend is rather save.
Considering the observational errors, such a small difference in the relative weights
could be in agreement with non-variability of the spectra.

In order to better understand the results,
the  distribution of the fractional abundance of some significant ions,
as well as the distribution of the temperature and density throughout the
SD and RD nebulae are given in Fig. 1 and Fig. 2, respectively.

The high relative weight of the RD model
indicates that a large region of gas, affected by photoionization from the
hot star and corresponding to characteristics closer to the ISM,
contributes to the observed fluxes.
The extended region  shows, in fact, relative small densities
and velocities not exactly fitting the conditions of the jet bulk.

 The increasing weight of the RD model with time could explain the decline of the SW jet.
In fact, more IS matter is swept up by the shock at latest epochs.  A larger contribution
corresponds to weaker fluxes because the RD model is characterized by relatively low densities
and velocities.
To obtain the distance r of the SW jet from the hot source, recall that the  fluxes
are calculated at the nebula. Adopting a distance d =
 200 pc, the data from Meier \& Kafatos (1995, Table 2C), and the results of model
m1$_{SW}$, we compare the observed and calculated fluxes of the NV line,
 1.5 10$^{-13}$ $\times$ (200 pc)$^2$ = 38.9$\times$ 0.0037 $\times$ r$^2$.
The distance
of the bulk of the SW jet from the center of the system   results r$\sim$ 6 10$^{14}$ cm.

\subsection{The NE jet}

The NE jet  is outlined in Meier \& Kafatos (1995, Fig. 1) and encompasses
the regions A, B, and D in VLA 6 cm observations.

\subsubsection{The optical line ratios}

The observed optical lines are from Hollis et al. (1991, Table 3)
and sample the regions near the peak of the jet emission.

The optical spectrum (Table 2) is explained by a composite model (m2$_{NE}$)
characterized by a  shock of 140 \kms, \Ts=80 000 K, and a rather low U=2.2 10$^{-3}$.
 Some line ratios (HeI/\Hb, [OIII]/\Hb,
[OI]/\Hb, ~and [SIII]/\Hb)  are better fitted
by a similar model with \Vs=150 \kms (n3$_{NE}$).

A low U, reduced by dilution,  confirms that the NE jet is rather far away
from the hot star. The distance r of the emitting nebula in the jet from
the  hot star is calculated relating the photon flux from the hot star,
$\cal N$ with the ionization parameter :
$\cal N$ (R$_{hot}$/r)$^2$ = U n c, where n is the density and
R$_{hot}$ = 5 10$^8$ cm is the radius of the hot star.
Adopting n = 3 10$^4$  \cm3 from Fig. 3, r results 2 10$^{15}$ cm.

\subsubsection{The evolution of the UV spectra}

Table 3 shows   the UV spectra  calculated by models m1$_{NE}$, m2$_{NE}$, m3$_{NE}$,
 and m4$_{NE}$ which are selected
to fit the spectra  of the NE jet in different epochs.
Model m4$_{NE}$,
which is characterized by conditions closer to those of the ISM,
is similar to m2$_{SW}$ (Table 1, column 3) that
was found to improve the fit of the SW jet spectrum
(relatively low \Vs, \n0, \B0, and a large D).
The models are all RD dominated. The line ratios refer to \Hb = 1.
In order to understand the line ratios, we present in Fig. 3 the distribution of the
fractional abundances of the most significant ions, as well as of the temperature and
density throughout the nebula corresponding to model m2$_{NE}$.

In Table 5 we compare model results with spectra at different epochs
for the NE jet.
The data (Meier \& Kafatos 1995, Table 2B) represent the fluxes at
different epochs.
 We could not  use  line ratios to CIV, as in Table 4,
because the CIV line is saturated.
Viotti et al. (1987) in 1985 Dec 23-25 measured the CIV flux of 28.8 10$^{-13}$ \erg.
This value is intermediate between the 1983 flux  and  the 1986  lower limit  (Table 5).

We obtain an acceptable fit of the data  summing up  m2$_{NE}$
with  model m4$_{NE}$ (Table 1, last column).
 The spectra  resulting from the weighted sum of m1$_{NE}$, m2$_{NE}$, and m3$_{NE}$ with
model m4$_{NE}$  are  presented in  Table 5 (M1$_{NE}$, M2$_{NE}$, and M3$_{NE}$,
respectively).
The relative  weights  appear in the last row of Table 3.
The arrows indicate that SiIV and OIV] line fluxes are summed up.

The fit is acceptable, although
 Table 5 shows that SiII,  [OII], and MgII are lower than observed.
The calculated fluxes of the SiIV and SiIII]  lines are higher than observed and could indicate that Si
is partly locked into dust grains.
The anomalous flux of the HeII 3203 line can be due to the fact that it falls at the
IUE sensitivity limit.

The  errors  in observations and  in model calculations   do not permit
to distinguish between  the  physical conditions in the emitting gas
between December 1987 and June 1989.

As found for the optical lines, a model with \Vs=150 \kms (m3$_{NE}$)
explains the spectra at later
epochs. Therefore, in Table 5 this model is compared with both the spectra
at 1987 and 1989,  while the model with \Vs=140 \kms better
fits the conditions at 29/12/1986.

A sensible difference of line fluxes is noticed  between May 1983 and December 1986.
 Although  many different models
should be considered, we have adopted model m4$_{NE}$  with the same
weight as in 1986-1989  and model m1$_{NE}$  characterized by a lower velocity and a higher density
than m3$_{NE}$.
This suggests that the shock has accelerated with
time  propagating  outwards the system, throughout  the decreasing density  gradient
of the red giant atmosphere (Contini 1997).

The ionization parameter U increasing at later epochs  indicates that it  changes
with the Mira pulsation period, rather than  with dilution.
The jets are perpendicular to the
accretion disk and  to the orbital plane. Therefore, obscuration of the hot
source   is due to  dusty material ejected  by the Mira.
A  time interval of 3.6 years between May 1983 and December 1986 corresponds to
3.4 pulsation periods, leading to a different U.

We can now calculate the distance  r of the NE jet from the binary system in the
different epochs.
Comparing
the observed NV 1240 flux at Earth in 1987 (15.2 10$^{-13}$ \erg)  with the flux calculated
by model m3$_{NE}$,  r  results  2   10$^{15}$ cm and in 1986  r = 1.8 10$^{15}$ cm.
In 1983 the flux of NV was 8.6 10$^{-13}$ \erg  and r=1.4 10$^{15}$ cm, indicating
that the NE jet has slightly farthened from the central system between 1983 and 1987.

To calculate the distance of the low density region  we consider the HeII 1640 line  in m4$_{NE}$
and find r=  2.26  10$^{15}$ cm.
The results confirm that different conditions coexist in the complex region
representing the jets.

In conclusion, the  observed trend of  increasing line fluxes with time is explained
by  a  slight acceleration of the  shock  expanding in the surrounding of the
binary system.

\begin{table*}
\centerline{Table 5}
\centerline{UV line  fluxes in the NE jet (10$^{-13}$ \erg)}
\begin{tabular}{ ll l l l l l ll } \\ \hline
    &25/5/1983&M1$_{NE}$ &29/12/1986 &M2$_{NE}$ &31/12/1987 &M3$_{NE}$ &5/6/1989&   \\
\hline
\ NV 1238+42 &8.6        &6.0&11.5     &12.0&15.2           &14.0& 14.9       \\
\ CII 1334+36&7.9        &10.4& 13.8    &21.6  & 14.7    & 22.7  & 16.5       \\
\ SiIV 1394 & 5.4      & 13.4 &9.1      &25.7 &10.1     &31.8  & 11.7     \\
\ OIV] 1401+07 &8.1     &$\uparrow$ &11.9       &$\uparrow$ & 15.4      &$\uparrow$ & 16.6&      \\
\ NIV] 1483+86&-   &5.0 &9.8        &7.12& 10.0        &8.3& 9.9      \\
\ CIV 1548+51 &14.6& 19.4   &$>$30.8 &31.0& $>$39.6  &33.5& $>$40.5   \\
\ HeII 1640&4.0         &3.4& 8.0      &7.94 &10.0      &8.0 & 9.5         \\
\ OIII] 1661+66 &7.2 & 5.0    &13.2     &10.5&15.0    &11.0&16.5     \\
\ NIII] 1749+54 &6.2& 5.0    &10.6& 10.0    &11.7& 10.5   &13.2    \\
\ SiII 1808+17 &-& 0.7       &4.4&  1.3   &5.5 & 1.6    & 7.2       \\
\ SiIII] 1883+92 &5.5 & 12.  &12.3  &  26.4      & 15.5 &27.2    & 15.7      \\
\ CIII] 1907+10 &$>$15.7&23.    & $>$27.5& 51.1    & $>$30.4& 53.7      &$>$33.8      \\
\ [OIII]+CII] 2320+& 13.2&  20.     & 38.4& 37.2    & 51.7& 41.5      &51.2      \\
\ [OII] 2470&8.0 & 5.0    & 14.4 & 7.0 & 20.3   &9.1  & 20.2      \\
\ MgII 2795+& 14.2      &6.0  & 18.5         & 12.7 &25.9 &13.7& 23.9     \\
\ HeII 3203 & - & 0.18    &   -& 0.5&      8.1  &0.4&6.6              \\            
\hline\\
\end{tabular}

\flushleft

$^1$ The observed data are in 10$^{-13}$ \erg

\end{table*}

\section{The HII region}

\begin{table*}
\centerline{Table 6}
\centerline{UV line intensities relative to  CIV (HII region)}
\begin{tabular}{ ll l l l l l ll } \\ \hline
     &8/11/1980 &m1$_{HII}$&25/5/1983&m2$_{HII}$& 25/10/1986 &m3$_{HII}$&26/5/1991&m4$_{HII}$ \\
\hline
\ NV 1238+42 &- &0.11    &- &0.11  &   0.27&0.26   & 0.20&0.22  \\
\ CII 1334+36&0.33 &0.30  & 0.29&0.29 & 0.27&0.24  & 0.28& 0.23 \\
\ SiIV 1394 & 0.22&0.1   &0.21&0.1  &0.23&0.1   & 0.27&0.1 \\
\ OIV] 1401+07 &0.32&0.32& 0.30&0.31  & 0.29&0.26  & 0.27&0.24 \\
\ NIV] 1483+86&-&0.2   &0.26?&0.2  & 0.27&0.22   & 0.2&0.22 \\
\ CIV 1548+51 &1.&1.&1.& 1.&1.&1.&1.&1.\\
\ HeII 1640&0.33&0.33   & 0.26&0.29 &0.2&0.18   & 0.18&0.174  \\
\ OIII] 1661+66 &0.55&0.33&0.45&0.29&0.35&0.2& 0.32&0.2\\
\ NIII] 1749+54 &0.5&0.4 &0.42&0.33&0.35&0.22&0.28&0.2\\
\ SiII 1808+17 &0.28&0.11&0.18&0.12&0.09&0.08& 0.14&0.08 \\
\ SiIII] 1883+92 &$>$0.63&1.1&0.79&0.88 & 0.54&0.5 & 0.38&0.45 \\
\ CIII] 1907+10 &$>$1.2&2.9 & $>$1.66&2.6 & $>$1.18&1.48 & $>$0.95&1.4 \\
\ [OIII]+CII] 2320+& $>$1.2&1.3 & 1.26&1.34&  0.67 &1.12& 0.9&1.08 \\
\ [OII] 2470&0.7&0.8  & 0.74&0.87 & 0.4&0.66 & 0.43&0.64 \\
\ MgII 2795+& $>$1.17&0.3 & 1.8&0.32 & 0.67&0.25 & 0.6&0.24 \\
\ HeII 3203 & - &0.02&   -&0.017 & 0.39&0.011 & 0.27&0.01 \\
\ \Hb  (\erg) &- & 4.86& - & 4.5 & - & 3.29& - & 3.02\\
\ CIV/\Hb & -& 1. & -&1. &-& 1.56 & - &1.6\\
\hline
\ \Vs (\kms)&-& 110 & - &110 & - &125 &  -&120\\
\ \n0 (\cm3)&-& 6(4)&-&6(4)&-&6(4)  &-&6(4) \\
\ \B0 (gauss)&-&2(-3)&-&2(-3)& - & 2(-3) & - & 2(-3)\\
\ D (10$^{14}$)& -&2. &-&2.&-& 2.& -& 2. \\
\ \Ts (K)&-&8(4)  &-&8(4)&- & 8(4) & - &8(4)\\
\  U &-&1(-2)&-&8.5(-3)&-&6(-3)&- &5.5(-3)\\
\hline\\
\end{tabular}

\end{table*}

We present in Table 6 the best fitting model of the spectra emitted
by the HII region in the different epochs. The data are taken from
Meier \& Kafatos (1995, Table 2A).
The  position of the IUE spectrograph aperture integrates
both the features identified as part of the HII region, as well as
the SW jet region (Meier \& Kafatos).
However, the fluxes emitted from the jets are by a factor $\geq$ 3 lower than those
from the HII region, so we will refer to these data as representative of  the HII region,
with a minor contamination  of the jets.

The models are described in the bottom of Table 6.
For  all  epochs a preshock density
of 6 10$^4$ \cm3  and \Vs ~slightly  higher than 100 \kms   explains almost all of the line ratios.
These conditions correspond to a postshock density
of $\geq$ 6. 10$^5$ \cm3. Notice, however, that the magnetic field
also plays an important role in downstream compression.
We have found \B0=2 10$^{-3}$ gauss, which is rather high, thus preventing
a large compression. A geometrical thickness,  D = 2 10$^{14}$ cm leads to the best agreement
of calculated with observed spectra.

The models adopted in Table 6 (m1$_{HII}$, m2$_{HII}$, m3$_{HII}$, and m4$_{HII}$) are calculated
considering that the shock and the photoionising
flux act on the same edge of the emitting nebula.

This condition corresponds to the reverse shock, which was
created by collision of the winds between the stars (Sect. 2).

The distribution of the fractional abundance of some significant ions, as well
as of the temperature and density downstream of the reverse shock are given in Fig. 4,

 Table 6 shows that NV/CIV slightly  changes with time, while  HeII/CIV decreases.
The results of model calculations generally show that NV/CIV increases with  \Vs ~,
while HeII/CIV and CIII]/CIV increase with both \Ts ~and U.
The ionization parameter is diluted by distance,  suggesting, therefore,
that the shock front is slightly farthening from the hot source.
The distance from the hot source in 1991, calculated by comparing the HeII line is 8.8 10$^{13}$ cm,
 while in 1980 the distance is 8.6 10$^{13}$ cm. The difference is  negligible, considering the errors
in the data and in the models.
Therefore, the reverse shock is most probably a standing shock.

The agreement of model results with the data is  good enough and indicates
that  the contribution of the expanding shock is negligible.
In fact, the expanding shock,  many years after  collision, has   reached the outskirts
of the system.
Here, the density is relatively low (10$^3$ \cm3) and the  nebula downstream of the
expanding shock is reached by a weak  radiation flux.
Therefore, the intensity of the lines is  by at least a factor of 10$^4$ lower
than the intensity of the lines emitted downstream of the reverse shock.

\section{The SED of the continuum}

\begin{figure}
\includegraphics[width=84mm]{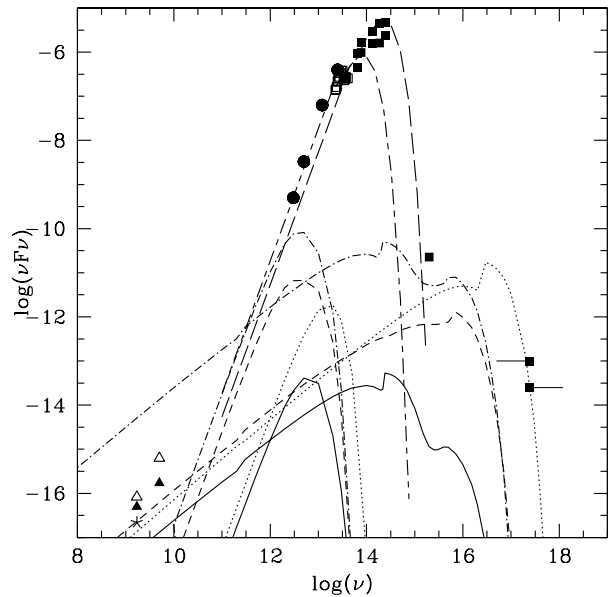}
\caption
{The comparison of the  observed continuum SED with model results (see text).
Solid lines represent the HII region (m4$_{HII}$); short-dashed lines
refer to the SW jet (m1$_{SW}$), dash-dotted lines to the NE jet
(m2$_{NE}$), and dotted lines to the model calculated with \Vs=300 \kms.}
\end{figure}

In Fig. 5 we compare the observed continuum SED  with  the  calculated  SEDs of the models
corresponding to the prevailing conditions in the R Aqr system.

The data in the near-IR (filled squares) come from Le Bertre (1993), those in the mid-IR (open squares) from
Monnier, Geballe, \& Danchi ( 1998),  in the far-IR are taken from IRAS (filled circles).
The data in the radio range come from Dougherty et al (1995).
The fluxes from the different features corresponding to  the central region (open triangles)
  and to the jets (filled triangles) have been summed up.

A  black-body radiation with a temperature of T=2800 K (long-dashed line)
represents   the red giant (Anandarao \& Pottasch 1986, Meier \& Kafatos 1995).
Moreover, we have added a black-body emission corresponding to T=1000 K (long-dash-short-dash line)
to represent emission from dust in the inner radius of the Mira (see Tuthill et al 2000).
Absorption by silicates at $\sim$ 9.7 $\mu$m is clearly noticed.

The models were selected by modelling the line spectra in previous sections.
Each model corresponds to two curves, one is bremsstrahlung radiation from the gas
and the other is reradiation by dust in the IR.
Dust grains and gas are coupled entering the shock front and mutually heat each other
(Viegas \& Contini 1994).
So, high velocity models correspond to mid-IR emission and low velocity models
to far-IR. In the present case,  there are no data in the very far-IR
to constrain the  dust-to-gas ratio (d/g = 4. 10$^{-4}$  is used in the models)
in the jet regions. However, Fig. 5 shows that
most of the corresponding frequency range is occulted by the black-body
emission from the red giant.

The models are  represented in Fig. 5 according to the relative weights.
They  are constrained only  by the datum in the UV (from Kafatos, Michalitsianos, \&
Hollis 1986) and by the data in the radio.
Model results should be summed up to  best fit the UV datum, however,
they are shown   one by one
in order to better understand single model contributions in the different frequency
ranges.
Solid lines  refer to model  M4$_{HII}$ which represents  the HII region and corresponds
to a shock velocity of \Vs = 120 \kms.
Short-dashed lines correspond to the SD model m1$_{SW}$,
dash-dotted lines correspond to model m2$_{NE}$.

Dotted lines  in Fig. 5 refer to  the model calculated with \Vs=300 \kms and \n0=1000 \cm3
(Michalitsianos et al. 1994).
 Relatively high velocities can survive when high velocity gas in the jets  collides with
ISM inhomogeneities  with a relatively low density. A shock velocity of 300 \kms
corresponds to a   temperature  of  1.4 10$^6$ K in the post shock region.
This temperature is responsible for soft X-ray emission
observed by Chandra  below 1.5 KeV,
while the hard X-ray emission observed from the central region of the system
comes from the accretion disk (Kellogg, Pedelty, and Lyon 2001).

The optical and UV spectra, however,
calculated with \Vs=300 \kms and \n0 $<$ 1000 \cm3

are weak compared with the spectra calculated for the jets and the HII region (see Sects. 3 and 4)
 because a) the gas  at a temperature of  $\sim$ 10$^6$ K will  hardly
recombine at a distance within the dimension of the system  and b) line intensities are
$\propto $ n$^2$. The \Hb ~flux  calculated with \n0 = 1000 \cm3 would be $\sim$ 4.3 10$^{-7}$ \erg
(cf. Table 1).
The corresponding NV and CIV lines  are weak but observable and may appear in the
wings of the line profiles as the contribution of a relatively broad component.
So,  this model  explains the soft X-ray data in the SED of the continuum,
but  its contribution to the line spectra is negligible.
This problem, however, awaits new higher resolution/higher sensitivity
space UV observations.

Fig. 5 shows that  the  radio data in the different positions follow a thermal curve
in agreement with Dougherty et al (1995).
Bremsstrahlung radiation from the NE jet overpredicts the data in the radio.
Recall that self-absorption increases with wavelength and can reduce
emission  at $\nu < $ 10 GHz (Contini 1997).

\section{Concluding remarks}

In the previous sections we have analysed the optical  spectra emitted from the jets and
the temporal evolution of the UV spectra from the  jets and  from the HII region.
We have demonstrated throughout the modelling of the different
regions of R Aqr that the coupled effect of radiation from the hot star
and shocks give a consistent explanation  to some of the
questions raised by the observational evidence.

The  modelling of the  spectra  indicates a  temperature of
\Ts=80 000  K  for the hot white dwarf component of the R Aqr system.
This high temperature, already suggested by Meier \& Kafatos,
is justified by the consistent account for the  shocks.

The time evolution of the NE jet is thus  explained by the decreasing of
the density gradient in the
outer region of the system and by a small change in the obscuration of the hot source
 by the Mira dusty envelope.
The spectral evolution of the NE jet indicate that
between 29/12/1986 and 5/6/1989 the shock velocity has increased by a very small
factor, from 140 \kms to 150 \kms. At earlier epochs (25/5/1983)
the physical conditions of the emitting gas were slightly different, corresponding
to lower \Vs (120 \kms), a higher density by a factor of $\sim$ 1.8, and a lower U by a factor of
$\sim$ 1.4.

The decrease of the SW UV line intensities between 1989 and 1991 is due to the increased contribution
of  low density-velocity  interstellar matter  from the region of the jet at later
epochs.

In the HII region the evolution of the UV spectra from 8/11/1980 to 26/5/1991
indicates that the reverse shock is quite likely a standing shock.

The results obtained by modelling the line spectra are cross-checked by the fit
of the continuum SED. It is found that a black-body temperature of 2800 K
reproduces the radiation from the red giant and an additional component of black-body emission
 of 1000 K  is emitted by dust  from the  envelop of the red giant.

Model calculations confirm that the radio emission is of thermal origin, however,  the
contribution of synchrotron radiation, created by Fermi mechanism at the shock front,
cannot be excluded.

We found that the  bulk region  of the NE jet  emission is at a distance  of
about 2 10$^{15}$ cm from the  binary system in 1987,  and
for the SW jet  is at $\sim$ 6 10$^{14}$ cm.
The distance of the reverse shock from the hot
source  is about 9 10$^{13}$ cm.

The magnetic field varies from 2 10$^{-3}$ gauss in the  HII region to
a minimum of 10$^{-4}$ gauss in the extended region.
In the bulk of the jets we found \B0 = 10$^{-3}$ gauss.
Recall that turbulence created by Rayleigh-Taylor instabilities near the shock front may increase the
magnetic field by some orders (see Contini \& Prialnik 1997).

\section{Aknowledgments}

We are very grateful to the referee for many
interesting remarks that highly improved the presentation of the paper.

\label{lastpage}

\end{document}